\documentstyle{article}
\begin{document}
\newcommand{\beq}{\begin{equation}}
\newcommand{\eeq}{\end{equation}}
\newcommand{\beqn}{\begin{eqnarray}}
\newcommand{\eeqn}{\end{eqnarray}}
\newcommand{\dpf}{\displaystyle\frac}
\newcommand{\no}{\nonumber}
\newcommand{\ep}{\epsilon}
\begin{center}
{\large Extreme black hole entropy obtained in an operational approach}
\end{center}
\vspace{1ex}
\centerline{\large Bin
Wang$^{a,b,}$\footnote[1]{e-mail:binwang@fma.if.usp.br},
\ Ru-Keng Su$^{c,}$\footnote[2]{e-mail:rksu@fudan.ac.cn}
and Elcio Abdalla$^{a,}$\footnote[3]{e-mail:eabdalla@fma.if.usp.br}
}
\begin{center}
{$^{a}$ Instituto De Fisica, Universidade De Sao Paulo,
C.P.66.318, CEP
05315-970, Sao Paulo, Brazil \\
$^{b}$ Department of Physics, Shanghai Teachers' University,
P. R. China
\\
$^{c}$ Department of Physics, Fudan University, Shanghai
200433, P. R.
China}
\end{center}
\vspace{6ex}
\begin{abstract}
The entropy of anti-de Sitter Reissner-Nordstr$\ddot{o}$m black hole is
found to be stored in the material which gathers to form it and equals 
to $A/4$ regardless of material states. Extending the study to two kinds
of extreme black holes, we find different entropy results for the first
kind of extreme black hole due to different material states. However 
for the second kind of extreme black hole the results of entropy are 
uniform independently of the material states. Relations between these 
results and the stability of two kinds of extreme black holes have 
been addressed.
\end{abstract}
\vspace{6ex}
\hspace*{0mm} PACS number(s): 04.70.Dy, 97.60.Lf
\vfill
\newpage

Traditionally it is widely believed that black holes have a gravitational
entropy given by $S_{BH}=A/4$, where $A$ is its area and units are such 
that $c=G=\hbar =k=1$. However until now the full understanding
of the origin of this entropy is still lacking, though some possible
explanations have been raised [1-3]. Recently, Pretorius, Vollick and
Israel have made significant progress on this problem [4]. By examining 
the
reversible contraction of a thin spherical shell down to the
Reissner-Nordstr$\ddot{o}$m (RN) black hole event horizon, they 
suggested 
that $S_{BH}$ is the equilibrium thermodynamic entropy that would be
stored in the material which gathers to form the black hole, if one
imagines all of this material compressed into a thin layer near its
gravitational radius. It is of interest to extend their study to other
black hole models and investigate whether their operational definition 
for
black hole entropy is valid for other black holes. This is the first
motivation of the present paper. In view of recent interest in anti-de
Sitter geometries, we will extend the study of ref. [4] to an
asymptotically
anti-de Sitter version of RN black hole [5].

The second motivation of this paper is to extend the operational
approach to extreme black hole (EBH) entropy. Recently there have been
heated discussions on EBH entropies and different results obtained by
using different treatments [6-10]. Starting with the original RN EBH,
Hawking et al claimed that a RN EBH has zero entropy, infinite proper
distance $l$ between the horizon and any fixed point [6,7]. However, in
the grand canonical ensemble, Zaslavskii argued that EBH can be
obtained 
as the limit of nonextreme counterpart by first adopting the boundary condition
$r_+=r_B$, where $r_+$ is the event horizon and $r_B$ is boundary of the cavity,
and then the extreme condition. The final extreme hole is in the 
topological
sector of nonextreme configuration and its entropy still obeys the
Bekenstein-Hawking formula [8-10]. 
Recently by using these two treatments, the
geometry and intrinsic thermodynamics have been investigated in detail
for
a wide class of EBHs including 4D and two-dimensional (2D) cases [11-13].
It was
found that these different treatments lead to two different
topological objects represented by different Euler characteristics and
show drastically different intrinsic thermodynamical properties both
classically and quantum-mechanically. Based upon these results it was
suggested that there maybe two kinds of EBHs in nature: the first
kind
suggested by Hawking et al with the extreme topology and zero entropy,
which can only be formed by pair creation in the early universe; on the
other hand,
the second kind, suggested by Zaslavskii, has the topology of
the nonextreme sector
and the entropy is still described by the Bekenstein-Hawking formula,
which can be developed from its nonextreme counterpart through second
order phase transition [11-13]. This speculation has been further
confirmed recently in a Hamiltonian framework [14] and using the grand
canonical
ensemble [15] as well as canonical ensemble [16] formulation for RN
anti-de Sitter black hole by finding that the Bekenstein-Hawking entropy
and
zero entropy emerge for extreme cases respectively. 
It is worth extending the operational approach [4] 
to
investigate the EBH entropy, especially for these two kinds of EBHs, 
and
compare the operational definitions of EBH entropy to the results
available. In [4] some attempts have been made to study the EBH entropy
and some ambiguous results have been obtained, however their study is 
only
limited in the first kind of RN EBH. By extending the study to two 
kinds
of RN EBH as well as anti-de Sitter RN EBH, we will show that ambiguous
results for EBH entropy do appear for the first kind of EBH, 
disappearing for the second kind of EBH. Some physical understanding on 
this
problem will be given.

The RN black hole solution of Einstein equations in free space with a
negative cosmological constant $\Lambda=-\dpf{3}{l^2}$ is given by [5]
\beq                    
{\rm d}s^2=-h{\rm d}t^2+h^{-1}{\rm d}r^2+r^2{\rm d}\Omega^2
\eeq
where
\beq             
h=1-\dpf{r_+}{r}-\dpf{r_+^3}{l^2r}-\dpf{Q^2}{r_+
r}+\dpf{Q^2}{r^2}+\dpf{r^2}{l^2}
\eeq
The asymptotic form of this spacetime is anti-de Sitter. There is an 
outer
horizon located at $r=r_+$. The mass of the black hole is given by
\beq           
m=\dpf{1}{2}(r_+ +\dpf{r_+^3}{l^2}+\dpf{Q^2}{r_+})
\eeq
In the extreme case $r_+, Q$ satisfy the relation
\beq               
1-\dpf{Q^2}{r_+^2}+\dpf{3r_+^2}{l^2}=0.
\eeq

Now we consider compressing a spherical shell reversibly from an 
infinite
radius down to the black hole event horizon.  We will concentrate our
attention on the NEBH case at the beginning with
the left-hand-side of Eq.(4) bigger than zero.
To maintain reversibility 
at  
each stage the shell must be in equilibrium with the acceleration
radiation that would be measured by an observer on the shell. To ensure
this equilibrium, an ``adiabatic" diaphragm must be interposed between 
the
faces. As done in [4], we picture the shell as a pair of concentric
spherical plates, with inner and outer masses $M_1$ and $M_2$, 
separated
by a massless and thermally inert interstitial layer of negligible
thickness. These two plates separate three concentric spherical 
regions:
an inner
region where $h(r)=h_1(r)$, a very thin intermediate flat region with
$h(r)=1$ and an outer region where $h(r)=h_2(r)$. The local temperature
$T_i$ of the plates is given by the expression
\beq                      
T_i=\dpf{h_i'}{4\pi\sqrt{h_i}}    \  (i=1,2)
\eeq

Introducing the Gaussian normal coordinates near every point on the 
plate
$\Sigma$, the coordinates on $\Sigma$ are $(\tau,\theta,\phi)$, where
$\tau$ is the proper time for an observer on $\Sigma$. Defining
$\vec{N}$ as the unit spacelike vector orthogonal to $\Sigma$ and
$\vec{U}$ the velocity of a mass element of this surface, the 
orthogonal
condition becomes $\vec{N}\cdot\vec{U}=0$. The velocity is
$\vec{U}=\dot{t}\partial_t+\dot{r}\partial_r$ where the overdot denotes
differentiation with respect to $\tau$. We obtain $\vec{N}=(\vert
g_{tt}\vert)^{-1}\dot{r}\partial_t +\vert
g_{tt}\vert\dot{t}\partial_r$. The normalization conditions are
$\vec{N}\cdot\vec{N}=1, \vec{U}\cdot\vec{U}=-1$. The extrinsic 
curvatures
relative to the Gaussian normal coordinates are simply
$K_{\tau\tau}=N_{\tau;\tau}=U^{\mu}U^{\nu}N_{\mu\nu}$ and $
              K_{xy}=N_{x;y} (x,y=\theta,\phi)$. Evaluating the jump
$\gamma^i_j$ in the extrinsic curvatures between different regions and
employing Israel's equation [17]
\beq                     
\gamma^i_j-\delta^i_j{\rm Tr}\gamma_{ij}=-8\pi s^i_j,
\eeq
we can obtain masses $M_i$ and surface pressures $P_i (i=1,2)$ for the
inner and outer plates,
\beqn        
M_i & = & R\xi_i(1-\sqrt{h_i(R)}) \\ \no
16\pi P_i & = & (\dpf{\xi_i h'_i(R)}{\sqrt{h_i(R)}}-\dpf{2M_i}{R^2})
\eeqn
where $R$ is the common radius of the two plates and $\xi_i=(-1)^i$.

The shell serves merely as the working substance and the nature of the
material in the shell is irrelevant, provided that the first law of
thermodynamics, 
\beq                
{\rm d}S=\beta{\rm d}M+\beta P{\rm d}A-\alpha{\rm d}N,
\eeq
is satisfied (we drop the index $i$ for the moment). Above, $\beta=1/T,
\alpha=\mu/T$ and $\mu$ is the chemical potential. $N$ here is 
introduced
as the number of particles in the shell to make the representation of
differential ${\rm d}S$ complete [4]. Using the Gibbs-Duhem relation
\beq                
S=\beta(M+PA)-\alpha N
\eeq
we have
\beq                 
n{\rm d}\alpha=\beta{\rm d}P+(\sigma+P){\rm d}\beta
\eeq
where $n=N/A$ and $\sigma$ is the surface mass density which satisfies
$\sigma=\dpf{M}{4\pi R^2}$. Substituting Eqs(5,7) into (10), we get 
\beq        
n{\rm d}\alpha=\sigma^2{\rm d}(\dpf{2\pi\sqrt{h}}{\sigma h'})
\eeq
The functions $n$ and $\alpha$ can be chosen arbitrarily if they satisfy
(11). 
For
simplicity, we can choose plate materials fulfilling the state equation
\beq               
n^*_i=\sigma^2_i\; ,\quad {\rm and}\; ,\quad \alpha^*_i
=\dpf{2\pi\sqrt{h_i}}{\sigma_i h'_i}.
\eeq
Considering the chemical potential $\mu^*_i=T_i\alpha^*_i$, we have 
\beq                 
\mu^*_i n^*_i=\sigma_i/2
\eeq
Substituting the above into Eq(9), we arrive at the 
entropy density $s_i=S_i/A$ of the plates as
\beq               
s^*_i=\beta_i P_i+\beta_i\sigma_i-\alpha^*_i n^*_i=\beta_i
P_i+\dpf{\beta_i\sigma_i}{2}.
\eeq
Using Eq(7) for the surface pressure and the local temperature of the
plate Eq(5), we find
\beq             
s^*_2=1/4
\eeq
For the outer plate, when it reaches the black hole horizon, its 
entropy
is one
quarter of its area in Planck units.

This result is also valid choosing general functions $n, \alpha$
which satisfy (11). The most general way of proceeding along these lines 
is
\beq              
\alpha_i=g_i(\alpha^*_i),  n_i=n^*_i/g'_i(\alpha^*_i)
\eeq
and
\beq          
\dpf{\mu_i n_i}{\sigma_i}=\dpf{g_i}{2\alpha^*_ig'_i}.
\eeq
The most general expression for the entropy density of the plates is
\beq               
s_i=\dpf{1}{4}[\xi_i+\dpf{4\pi\sqrt{h_i}}{h'_i}(2\sigma_i -8\mu_i n_i)]
\eeq
When the outer plate approaches  the horizon, $h_2\rightarrow 0$ as
$R\rightarrow r_+$. Therefore $s_2=1/4$ again. This means that in the
anti-de Sitter RN NEBH, regardless of equations of states, the entropy 
of
a shell made of materials approaches $ A/4$ as the shell approaches its
event horizon. This result is in agreement with [4] for RN NEBH.

Now it is of interest to extend the above discussion to EBH cases. As
stated in [11-13], two kinds of EBHs emerge due to two different
treatments. The first kind of EBH obtained by Hawking et al is the
original EBH with zero entropy, zero Euler characteristic and arbitrary
imaginary time period $\beta$ because of its peculiar topology and no
conical singularity for its spacetime [6,7]. While the second kind of 
EBH
proposed by Zaslavskii has entropy equaling $A/4$ and the same 
topology
as that of NEBH [8-10]. We hope that using operational approach can 
give a deeper understanding of these two kinds of EBHs' entropies.

In [4], attempts have been given to find the RN EBH entropy by using
operational approach. However in their study, the authors only consider
the 
first
kind of RN EBH case with $\beta$ arbitrary and arrive at 
$s^*_{EBH}(1)=0$
and
$s_{EBH}(1)=\dpf{1}{2}(1-\dpf{\mu n}{\sigma})$ for different states of
materials in the extremally charged spherical shell collapsing onto the
hole. They concluded that entropy of RN EBH may depend on
their prior history. This result is not valid for the second kind 
of
RN EBH as we will show in the following. This kind of EBH, obtained by
first taking the
boundary limit and then the extreme limit in the grand canonical 
ensemble, has the same topology as that 
of NEBH,
and still has a conical singularity in the spacetime. Therefore, $\beta$
cannot be
arbitrary, being given by $1/T$, where $T$ is the
nonzero local
temperature [8].  We keep this fact in mind and let a
nonextreme shell collapse to black hole horizon first and make it 
become
extreme afterwards as we have done for obtaining the second kind of EBH.
We thus find, for the simplest choice of $n, \alpha$ as (22) in [4],
\beqn                 
s^*_{EBH}(2) & = & \beta_2 P_2       \\ \no
          & = & \dpf{4\pi
V_2}{f'_2}\dpf{1}{16\pi}(\dpf{\xi_2f'_2}{V_2}-\dpf{2M_2}{R^2})=1/4,
\eeqn
where we first took  the boundary limit $R\rightarrow r_+$,
which leads to
$V_2=\sqrt{f_2(R)}\rightarrow 0$.

However when $\alpha, n$ fulfill the general formulas (29,30) of [4], we
still have
\beqn        
s_{EBH}(2) & = & \beta_2 P_2+\beta_2\sigma_2(1-\dpf{\mu_2 
n_2}{\sigma_2})
\\ \no
          & = &
\dpf{4\pi 
V_2}{f'_2}\dpf{1}{16\pi}(\dpf{\xi_2f'_2}{V_2}-\dpf{2M_2}{R^2})+\dpf{4\pi
V_2}{f'_2}\sigma_2(1-\dpf{\mu_2 n_2}{\sigma_2})=1/4
\eeqn
in the boundary limit, taking  $R\rightarrow r_+$ and
$V_2\rightarrow 0$.

These results indicate that unlike the results obtained for the first 
kind of RN EBH, the operational approach leads to universal results 
for the entropy of the second kind of RN  EBH. These results also 
hold in Anti-de Sitter RN EBH.

For the first kind of  Anti-de Sitter RN EBH, because of their peculiar
topology, it
has been shown that its imaginary time period $\beta$ is arbitrary [5,18]. 
Taking
account of this fact, from Eq(14) and (7), we find for the extreme 
shell satisfying the equation of state (12)
\beq        
s^*_{EBH}(1)=\beta
\dpf{\xi_2}{16\pi}\dpf{h'_i}{\sqrt{h_i}}=\beta\dpf{\xi_2}{16\pi}2(\sqrt{h_i})'=0
\eeq
We took $R\rightarrow r_+ (h_i\rightarrow 0)$. It is the same as that  
in the
RN first kind of EBH case. However it is worthy to point out that the
entropy here decreases as
the extreme shell approaches  the black hole horizon, unlike the first
kind of RN EBH where
$s^*_{EBH}(1)=0$ at all stages. This is because in the first kind of RN
EBH, the surface pressure for the extremely charged shell material is
always zero, while this does not hold for the first kind of anti-de 
Sitter
RN EBH. 

For general state satisfying (16,17),
\beq                
s_{EBH}(1)=\dpf{\beta}{4}(2\sigma_i-8\mu_i n_i)
\eeq
when the extreme shell collapse to the black hole horizon. These 
results
again support the
argument that the entropy of the first kind of EBHs may depend on their 
prior
history [4].

Now we turn to study entropy for the second kind of anti-de Sitter RN 
EBH. It has been shown that it has
the same topology as that of the NEBH, therefore it has a conical 
singularity with
$\beta=1/T$ [15,18]. $T$ here is the local temperature
$T=T_H/[h(r_B)]^{1/2}$ and $T_H$ is the Hawking temperature. For the
second kind of EBH $T$ is nonzero though $T_H =0$ [18]. And in the 
grand canonical ensemble actually only the local temperature $T$ has
physical meaning, whereas $T_H$ can always be rescaled without changing 
observable quantities [19]. Therefore, this kind of EBH can be achieved
with no 
contradiction with the third law of thermodynamics. Let the nonextreme
shell collapse to the
black hole and make
it extreme afterwards,  which corresponds to the treatment of Zaslavskii 
by
first adopting
the boundary limit and then the extreme limit. We have 
\beq               
s^*_{EBH}(2)=\{\dpf{4\pi\sqrt{h_2}}{h'_2}[\dpf{\xi_2
h'_2}{16\pi\sqrt{h_2}}]\vert_{R\rightarrow r_+}\}\vert_{extr}=1/4
\eeq
for the equation of state of the shell material given by (12).

For the shell material satisfying the general equations of (16,17) 
\beq                
s_{EBH}(2)=\{\dpf{4\pi\sqrt{h_2}}{h'_2}[\dpf{\xi_2
h'_2}{16\pi\sqrt{h_2}}+\dpf{\sigma_2}{4}(2-\dpf{8\mu_2
n_2}{\sigma_2})]_{R\rightarrow r_+}\}\vert_{extr}=1/4,
\eeq
by taking $R\rightarrow r_+$ first ($h_2\rightarrow 0$ first).
Therefore the entropies of this second kind of EBH are independent of  
their
prior
history as that of the second kind of RN EBH case. 

At first sight, it seems hard to believe why the operational approach 
leads to
drastically different results for two kinds of EBHs. Especially the 
changable
results for entropy of the first kind of EBH concerning different 
equations of state
of collapsing materials. We know
that entropy is a function of state. We have shown in our previous 
papers [11-13,18]
that although both of these two kinds of EBHs satisfy the extreme 
condition, their
topological properties differ drastically, so we cannot treat them as 
the same
state. This understanding may help us to understand the different 
entropies for two
kinds of EBHs. But how can we explain  different operational 
definitations
of the
entropy for the first kind of EBH?

As an example, let us first go over the issue of stability discussed for 
RN black hole [20]. The heat capacity at constant electrostatic potential
difference and cavity radius can be computed from $C_{\phi,
r_B}=-\beta(\dpf{\partial S}{\partial \beta})_{\phi, r_B}$, which leads
to  $C=4\pi r_B^2
x^3(1-x)/(3x^2-2x-q^2),$
where $x=r_+/r_B, q=e/r_B$ Eq(5.17) of [20]. For the first kind of EBH
obtained by 
Hawking et al's
treatment (adopting extreme condition at the very beginning) $C=4\pi 
r_B^2
x^3(1-x)/2x(x-1)<0$. The negative sign here determine that this kind of 
EBH is locally
unstable. In addition, it follows from Eq(5.9-5.11) of [20], that the
second
derivatives of the action $I$ with respect to  entropy $S$ and the mean
charge value
$<Q>$ diverge by starting with the original EBH [for example,
$\partial^2 I/\partial<Q>^2=4\pi(1-x)(1-q^2/x^2)^{-1}(1-q^2/x)^{-1}, 
q=x$ for the
original EBH]. This means that fluctuations of the charge and entropy 
are infinite
for the first kind of EBH. Because of this instability of the first 
kind of EBH,
different states of materials collapses onto it will of course lead to 
different
entropy results, while for the second kind of EBH, the entropy is 
uniform because
of its stability, as shown in [8]. The extension of the stability 
results
to two kinds of anti-de Sitter
RN EBHs is obvious. 

In summary, we have extended the operational definition of RN black 
hole entropy [4] to an interesting anti-de Sitter RN NEBH model and found
that the entropy for this NEBH is described by the Bekenstein-Hawking 
formula as well. Extending the operational approach to two kinds
of EBHs we suggested [11-13], we arrived at different results for their
entropy. For the first kind of EBH, the entropy values depends on 
different material equations of state which collapse onto the hole, 
which is in agreement with the results of [4]. However, for the second 
kind of EBH, uniform entropy emerges, regardless of the material 
equation of state. These different results can be attributed to the 
issues of stability for these two kinds of EBHs. Using the argument 
given in [20], it is easy to find that the first kind of EBH is
unstable. Thus its entropy changes in case of different states of the
material collapsing onto it. However the second kind of EBH is stable
[8], and it has the same values of entropy, regardless of material states.  
Therefore, although two kinds of EBHs can be created, due to stability
only the second kind of EBH can last for long in  nature. 

It is worth pointing out that all discussions for obtaining EBH in this
paper are just from a theoretical viewpoint and refers to the mathematical
treatment only. Physical realization for creating EBHs leads to the
problem about how to satisfy the third law of thermodynamics. Although 
it is clear that quantum processes like evaporation, which typically 
involve the absorption of negative energy, can violate  Nernst's form of 
the third law, in all classical studies  it holds [4,21-23].
The fact that the entropy of the first kind of EBH tends, as $T\rightarrow 0$,
to an absolute zero, ensures that the strong version of the third law holds
in this kind of EBH. For the second kind of EBH obtained by first letting the
nonextreme shell collapse to the black hole horizon and next making it become
extreme does not upset the third law as well, because the final extreme
state can be obtained at nonzero temperature. Therefore no challenge to 
the third law arise. Besides mathematical treatments, physical processes 
for obtaining EBH still need further studies. This is still an open 
question and we shall discuss it elsewhere.

ACKNOWLEDGMENT: 
This work was partically supported by Fundac\~{a}o de Amparo \`{a} 
Pesquisa do Estado de S\~{a}o Paulo (FAPESP) and Conselho Nacional de 
Desenvolvimento Cient\'{\i}fico e Tecnol\'{o}gico (CNPQ).
B. Wang would  like to acknowledge the support given by Shanghai Science
and Technology Commission.

\end{document}